\documentclass[11pt]{article}

\usepackage{times}
\usepackage{amssymb}
\usepackage{amsmath}
\usepackage{latexsym}
\usepackage{graphics}
\usepackage{color}
\usepackage{url}

\newtheorem{claim}{Claim}
\newtheorem{observation}{Observation}
\newtheorem{theorem}{Theorem}
\begin{document}

\title{On the complexity of finding a sun in a graph}

\date{\empty}

\author{
Ch\'inh T. Ho\`{a}ng\thanks{Department of Physics and Computer
Science, Wilfrid Laurier University, Canada.  Research supported
by NSERC. \texttt{ email: choang@wlu.ca}}  }

\maketitle
\begin{abstract}
The sun is the graph obtained from a cycle of length even and at
least six by adding edges to make the even-indexed vertices
pairwise adjacent. Suns play an important role in the study of
strongly chordal graphs.  A graph is  chordal if it does not
contain an induced cycle of length at least four. A graph is
strongly chordal if it is chordal and every even cycle has a chord
joining vertices whose distance on the cycle is odd. Farber proved
that a graph is strongly chordal if and only if it is chordal and
contains no induced suns. There are well known polynomial-time
algorithms for recognizing a sun in a chordal graph. Recently,
polynomial-time algorithms for finding a sun for a larger class of
graphs, the so-called HHD-free graphs, have been discovered.  In
this paper, we prove the problem of deciding whether an arbitrary
graph contains a sun in NP-complete.
\end{abstract}

\section{Introduction}
\begin{figure}
  \begin{center}
  \includegraphics{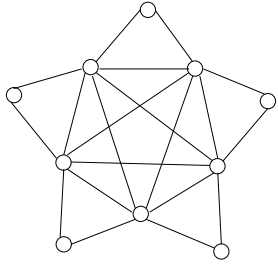}
  \end{center}
  \caption{The 5-sun}
  \label{fig:5sun}
\end{figure}

A {\it hole} is an induced cycle with at least four vertices. A
graph is {\it chordal} if it does not contain a hole as an induced
subgraph. Farber \cite{farber} defined a graph to be {\it strongly
chordal} if it is chordal and every cycle in the graph on $2k$
vertices, $k \geq 3$, has a chord $uv$ such that each segment of
the cycle from $u$ to $v$ has an odd number of edges. We denote by
{\it $k$-sun} the graph obtained from a cycle of length $2k$
($k\geq 3$)  by adding edges to make the even-indexed vertices
pairwise adjacent. Figure \ref{fig:5sun} shows a 5-sun. A {\it
sun} is simply a $k$-sun for some $k \geq 3$. Farber showed
\cite{farber} that a graph is strongly chordal if and only if it
is chordal and does not contain a sun as induced subgraph.
Farber's motivation was a polynomial-time algorithm for the
minimum weighted dominating set problem for strongly chordal
graphs. The problem is NP-hard for chordal graphs \cite{boojoh}.
In this paper, we prove that it is NP-hard to find a sun in an
arbitrary graph. This result is motivated by the following
discussion on chordal and strongly chordal graphs. For more
information on this topics, see \cite{bra, gol}.

\begin{figure}
  \begin{center}
  \resizebox{!}{1.3in}{\includegraphics{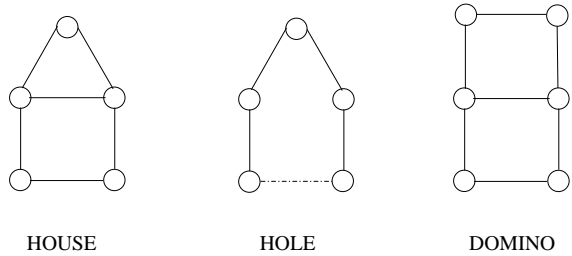}}
  \end{center}
  \caption{The house, the hole and the domino}
  \label{fig:HHD}
\end{figure}

We use $N(x)$ to denote the set of vertices adjacent to vertex $x$
in a graph $G$. Define $N[x]= N(x) \cup \{x\}$. A vertex $x$ in a
graph is {\it simplicial} if $N(x)$ induces a complete graph. It
is well known \cite{D} that graph $G$ is chordal if and only if
every induced subgraph $H$ of $G$ contains a simplicial vertex of
$H$. Farber proved \cite{farber} an analogous characterization for
strongly chordal graphs. A vertex $x$ in a graph is {\it simple}
if the vertices in $N(x)$ can be ordered as ${x_1},{x_2}, \ldots,
{x_k}$ such that $N[{x_1}] \subseteq N[{x_2}] \subseteq \ldots
\subseteq N[{x_k}]$. Thus, every simple vertex is simplicial. For
a graph $G$, let ${\cal R} = {v_1},{v_2}, \ldots, {v_n}$ be an
ordering of vertices of $G$. Let $G(i)~=~G[\{{v_i}, {v_{i+1}},
\ldots, {v_n}\}]$, i.e., the subgraph induced in $G$ by the set
$v_i$ through $v_n$ of vertices. ${\cal R}$ is a {\it simple
elimination ordering} for $G$ if $v_i$ is simple in $G(i)$, $1
\leq i \leq n$.  The following is due to Farber \cite{farber}:
\begin{theorem}[\cite{farber}]
The following are equivalent for any graph $G$:
\begin{itemize}
 \item $G$ is strongly chordal.
 \item $G$ is chordal and does not
contain a sun.
 \item Vertices of $G$ admit a simple elimination
ordering.
\end{itemize}
\end{theorem}
Thus, suns play an important role in the studies of chordal and
strongly chordal graphs. There are well known algorithms
\cite{tarpai, lubiw} to test whether a chordal graph is strongly
chordal and thus whether it contains a sun. It is natural to
investigate the problem  for larger classes of graphs. A graph is
HHD-free if it does not contain a house, a hole, or a domino (see
figure \ref{fig:HHD}). Every chordal graph is a HHD-free graph.
HHD-free graphs \cite{hoakho} have several properties analogous to
those of chordal graphs. Brandst\"adt \cite{problem} proposed the
problem of finding a sun in a HHD-free graph. This problem was
proved to be polynomial-time solvable in \cite{g2} and
\cite{eschoa}.
In this paper, we will prove the following
\begin{theorem}\label{thm:sun}
It is NP-complete to decide whether a graph contains a sun.
\end{theorem}
Denote by {\it k-hole} the hole on $k$ vertices. A {\it
k-antihole} is the complement of a $k$-hole. A graph is {\it
weakly chordal} \cite{hay} if it does not contain a $k$-hole or
$k$-antihole with $k \geq 5$. Weakly chordal graphs generalize
chordal graphs in a natural way, and they are known to be perfect
and have many interesting algorithmic properties (see
\cite{hayspi}). In spite of Theorem \ref{thm:sun}, it is
conceivable there are polynomial-time algorithms to solve the sun
recognition problem for weakly chordal graphs  or even perfect
graphs \cite{ramree}. In this spirit, we will refine Theorem
\ref{thm:sun} to obtain a stronger result.
\begin{theorem}\label{thm:sun2}
It is NP-complete to decide whether a graph $G$ contains a sun,
even when $G$ does not contain a $k$-antihole with $k \geq 7$.
\end{theorem}
Let $k$-CLIQUE (respectively, $k$-SUN) be the problem whose
instance is a graph $G$ and an integer $k$, for which the question
to be answered is whether $G$ contains a clique on $k$ vertices
(respectively, $k$-SUN). It is well known \cite{kar} that
$k$-CLIQUE is NP-complete. It is not difficult to prove, but
perhaps interesting to note that $k$-SUN is also NP-complete.
Observe that if $k$ is a constant (not part of the input), then
the two problems can obviously be solved in polynomial time.
\begin{theorem}\label{thm:ksun}
$k$-SUN is NP-complete.
\end{theorem}
Note that Theorem \ref{thm:sun} implies Theorem \ref{thm:ksun}: To
decide whether a graph contains a sun, we only need solve $O(n)$
instances of $k$-SUN with $k$ running from $3$ to $n/2$, where $n$
is the number of vertices of the graph. However, we have a short
and direct proof of Theorem \ref{thm:ksun}.  We will give the
proofs of Theorems \ref{thm:sun}, \ref{thm:sun2} and
\ref{thm:ksun} in the remainder of the paper.
\section{The proofs}
First, we need introduce some definitions. For simplicity, we will
say a vertex $x$ {\it sees} a vertex $y$ if $x$ is adjacent to
$y$; otherwise, we will say $x$ {\it misses} $y$. Let $G,F$ be two
vertex-disjoint graphs and let $x$ be a vertex of $G$. We say that
a graph $H$ is obtained from $G$ by {\it substituting} $F$ for $x$
if $H$ is obtained by replacing $x$ by $F$ in $G$ and adding the
edge $ab$ for any $a \in V(G) -\{x\}$, and any $b\in F$ whenever
$ax$ is an edge of $G$. In the proofs, we will often use the
observation that every vertex in $H-F$ either sees all, or misses
all, vertices of $F$.

By ($c_1, c_2, \ldots, c_k$, $e_1, e_2, \ldots, e_k$) we denote
the $k$-sun with vertices  $c_1, c_2, \ldots, c_k$, $e_1, e_2,
\ldots, e_k$ such that  $c_1, c_2, \ldots, c_k$ induce a clique,
$e_1, e_2, \ldots, e_k$ induce a stable set, each $e_i$ has degree
two and sees $c_i, c_{i+1}$ with the subscripts taken modulo $k$.
The vertices $e_i$ will be called the {\it ears} of the $k$-sun. A
{\it triangle} is a clique on three vertices.

We will rely on the following NP-complete problem  due to Poljak
\cite{pol}.

\noindent {\bf STABLE SET IN TRIANGLE-FREE GRAPHS}\\
\noindent
Instance: A triangle-free graph $G$, an integer $k$.\\
Question: Does $G$ contain a stable set with $k$ vertices?\\

\noindent {\it Proof of Theorem \ref{thm:sun}.} We will reduce
STABLE SET IN TRIANGLE-FREE GRAPHS to the problem of finding a sun
in a graph.

\begin{figure}
  \begin{center}
  \resizebox{!}{0.5in}{\includegraphics{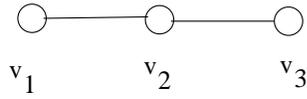}}
  \end{center}
  \caption{The graph G}
  \label{fig:p3}
\end{figure}

\begin{figure}
  \begin{center}
  \resizebox{!}{4in}{\includegraphics{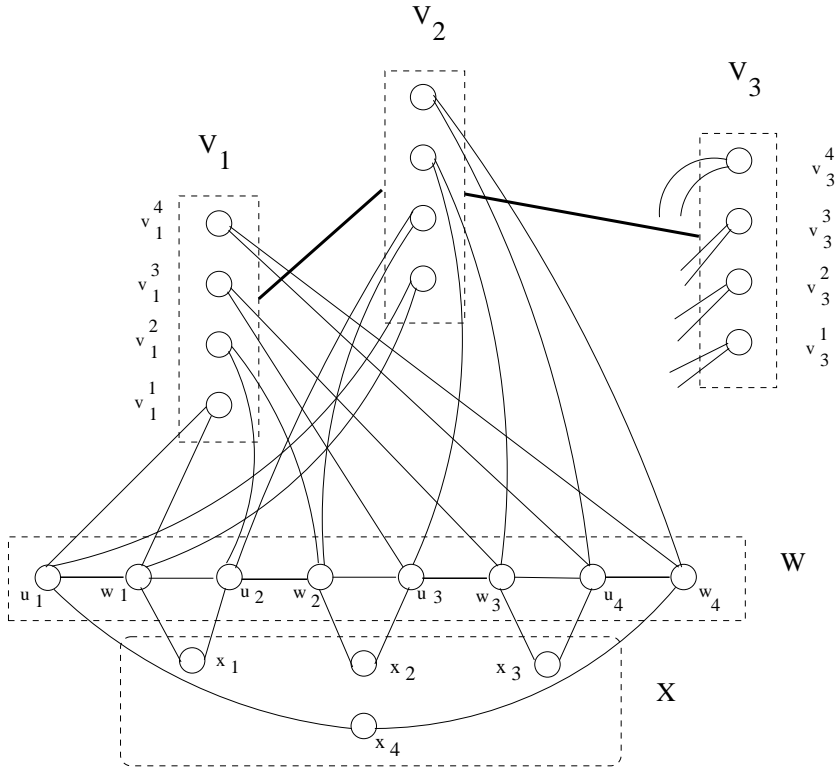}}
  \end{center}
  \caption{The graph f(G,4)}\label{fig:fG}
\end{figure}

Let $G=(V,E)$ be a triangle-free graph with $V = \{ v_1, v_2,
\ldots , v_n \}$, and without loss of generality assume $k\geq 4$.
Define a graph $f(G,k)$ from $G$ as follows. Substitute for each
vertex $v_i$  a clique $V_i = \{ v_i^1, v_i^2, \ldots, v_i^k \}$;
add a clique $W$ with vertices $u_1, w_1, \ldots, u_k, w_k$; add a
stable set $X$ with vertices $x_1, \ldots, x_k$; for $i=1,2,
\ldots , k$, add edges $x_i w_i$ and $x_i u_{i+1}$ (the subscripts
are taken module $k$); for $i = 1, 2, \ldots, n$ and $j= 1,2,
\ldots k$, add edges $v_i^j u_j, v_i^j w_j$. Figure \ref{fig:p3}
shows a graph $G$ whose graph $f(G,4)$ is shown in Figure
\ref{fig:fG} (for clarity, we do not show all edges of $f(G,4)$;
all adjacency between $V_1$ and $W$, and between $V_2$ and $W$ are
shown, adjacency between $V_3$ and $W$ are not shown; the thick
line between $V_1$ and $V_2$ (and between $V_2$ and $V_3$)
represents all possible edges between the two sets; there are no
edges between $V_1$ and $V_3$; each of the sets $V_i$, $W$ induces
a clique; the set $X$ induces a stable set.)
We will often rely on the following observations.
\begin{observation}\label{obs:triangle}
Suppose $G$ is triangle-free. Then $f(G,k)$ does not contain a
triangle each of whose vertices belongs to a distinct $V_i$.
$\Box$
\end{observation}
\begin{observation}\label{obs:VW}
Let $x$ be a vertex in $V_i$, $y$ be a vertex in $V_j$ with $i
\not= j$. If $x$ and $y$ have a common neighbor $z$ in $W$, then
$N(x) \cap W = N(y) \cap W$. $\Box$
\end{observation}
The theorem follows from the following claim.
\begin{claim}\label{cla:main}
\mbox{$G$ has a stable set with $k$ vertices if and only if
$f(G,k)$ contains a sun.}
\end{claim}
\noindent {\it Proof of Claim \ref{cla:main}.} Suppose $G$ has a
stable set with vertices $v_1, v_2 , \ldots , v_k$. Then $f(G,k)$
has a $2k$-sun $(c_1, c_2, \ldots , c_{2k}, $ $e_1, e_2, \ldots ,
e_{2k})$ with  $e_{2i - 1} = v_i^i $, $e_{2i} = x_i$,  $ c_{2i -1}
= u_i $, $c_{2i} = w_i $, for $i = 1, 2, \ldots k$.

Now, suppose $f(G,k)$ contains a sun. Write $U = V_1 \cup V_2 \cup
\ldots \cup V_n$.  We will establish that
\begin{eqnarray}
\mbox{Any sun $S$ of $f(G,k)$ is a $2k$-sun with $k$ ears in $V_1
\cup \ldots \cup V_n$.}\label{eq:sun}
\end{eqnarray}

Consider a sun $S = (c_1, c_2, \ldots , c_t, e_1, e_2, \ldots ,
e_t)$ of $f(G,k)$. First, we claim that (with the subscript  taken
modulo $k$)
\begin{eqnarray}
\mbox{If an ear $ e_j$ lies in $X$, then $e_{j-1}, e_j$ lie in
$U$}.\label{eq:X}
\end{eqnarray}
 Let
$x_i$ be a vertex in $X$ that is an ear $e_j$ of $S$. We may
assume that $c_j = w_i$ and $c_{j+1} = u_{i+1}$. Since $e_{j-1}$
sees $w_i$ and misses $u_{i+1}$, we have $e_{j-1} \in V_s$ for
some $s$. Similarly, we have $e_{j+1} \in V_r$ for some $r$. Note
that $r \not= s$. So, (\ref{eq:X}) holds.

Since $W$ is a clique, $S$ must have an ear in $U \cup X$.
(\ref{eq:X}) implies that
\begin{eqnarray}
\mbox{$U$ contains an ear of $S$.}\label{eq:ear}
\end{eqnarray}
Next, we will prove
\begin{eqnarray}
     \mbox{If $e_i \in V_j$ then $c_i, c_{i+1} \in W$}. \label{eq:inK}
\end{eqnarray}

Suppose (\ref{eq:inK}) is false.  For simplicity, we may assume
$i=1$ and $j=1$ (we can always rename the vertices of $f(G,k)$ and
$S$ so that this is the case). We will often implicitly use the
fact that a vertex in $V_a$ either sees all, or misses all,
vertices of $V_b$ whenever $a \not= b$. We will distinguish among
several cases.

{\it Case 1: $c_1 , c_2 \in V_1$}. Since $c_3$ sees $c_1 , c_2$
and misses $e_1$, $c_3$ cannot be in $U$. Thus, $c_3$ is in $W$.
But no vertex in $W$ can see two vertices in $V_1$, a
contradiction.

{\it Case 2: $c_1 \in V_1, c_2 \in V_j$ for some $j \not= 1$.} We
may write $j=2$. Since $c_3$ (respectively, $e_t$) sees $c_1$ and
misses $e_1$, $c_3$ (respectively, $e_t$) cannot be in $U$. Thus,
$c_3$ and $e_t$ are in $W$.  Observation \ref{obs:VW}, with
$z=c_3, x = c_1, y = c_2$, implies $e_t$ sees $c_2$, a
contradiction to the definition of $S$.

{\it Case 3: $c_1 \in V_1, c_2 \in W$.} This case is not possible
since a vertex in $W$ can have at most one neighbor in any $V_j$.

{\it Case 4: $c_1 , c_2 \in V_j$ for some $j \not= 1$.} We may
write $j = 2$. Since $e_2$ sees $c_2$ and misses $c_1$, $e_2$ is
in $W$. Since $e_t$ sees $c_1$ and misses $c_2$, $e_t$ is in $W$.
But then $e_2$ sees $e_t$, a contradiction.

{\it Case 5: $c_1 \in V_j, c_2 \in V_r$ with $j \not= r, j \not=
1, r \not= 1$.} In this case, $e_1, c_1, c_2$ contradict
Observation \ref{obs:triangle}.

{\it Case 6: $c_1 \in V_j$ for some $j \not= 1$ and $c_2 \in W$.}
We may let $j=2$. If $c_3 \in W$, then Observation \ref{obs:VW},
with $z=c_2, x=e_1, y=c_1$ implies $c_3$ sees $e_1$, a
contradiction to the definition of $S$. So,  we have $c_3 \in U$.
Since $c_3$ misses $e_1$, we have $c_3 \not\in V_1 \cup V_2$. So,
we may assume $c_3 \in V_3$. We have $e_2 \not\in W$; for
otherwise Observation \ref{obs:VW}, with $z=c_2 ,x = c_1, y =
c_3$, implies $e_2$  sees $c_1$, a contradiction to the definition
of $S$. We have $e_2 \not\in V_1 \cup V_2 \cup V_3$ since $ e_2$
misses $e_1$ and $c_1$. So, we may assume $e_2 \in V_4$. Since
$e_3$ (respectively, $c_4$, if it exists) sees $c_3$ and misses
$e_1$, Observation \ref{obs:VW}, with $z=c_2, x=e_1, y = c_3$,
implies $e_3 \not\in W$ (respectively, $c_4 \not\in W$). Since
$e_3$ (respectively, $c_4$, if it exists) misses $e_1$ and $e_2$,
we have $e_3 \not\in V_1 \cup V_2 \cup V_3 \cup V_4$
(respectively, $c_4 \not\in V_1 \cup V_2 \cup V_3 \cup V_4$). Now,
if $t=3$, then the three vertices $e_3, c_1, c_3$ contradict
Observation \ref{obs:triangle}. But if $t > 3$, then the three
vertices $c_4, c_1, c_3$ contradict Observation
\ref{obs:triangle}.

So (\ref{eq:inK}) holds. Next, we will establish two more
assertions (where the subscripts are taken modulo $k$) below.
\begin{eqnarray}
\mbox{If an ear $ e_j$ lies in $U$ then $e_{j-1}, e_{j+1}$ lie in
$X$}.\label{eq:V}
\end{eqnarray}
By (\ref{eq:inK}) and the definition of $f(G,k)$, we may assume
$c_i = u_i, c_{i+1} = w_i$. Since $x_i$ is the only vertex of
$f(G,k)$ that sees $w_i$ and misses $u_i$, we have $x_i =
e_{j+1}$. Similarly,  we have  $x_{i-1} = e_{j-1}$. So,
(\ref{eq:V}) holds.
\begin{eqnarray}
\mbox{If some vertex $x_i \in X$ is an ear of $S$, then $x_{i+1}$
is also an ear of $S$}.\label{eq:XX}
\end{eqnarray}
Let $x_i$ be a vertex in $X$ that is an ear $e_j$ of $S$.  We may
assume that $c_j = w_i$ and $c_{j+1} = u_{i+1}$. By (\ref{eq:X}),
we have $e_{j+1} \in V_a$ for some $a$. By (\ref{eq:inK}), we have
$c_{j+2} = w_{i+1}$. By (\ref{eq:V}), $e_{j+2}$ lies in $X$, and
so we have $e_{j+2} = x_{i+1}$. Thus, (\ref{eq:XX}) holds.

We are now in position to prove (\ref{eq:sun}). From
(\ref{eq:ear}),  we may assume $e_1$ lies in $U$. By (\ref{eq:V}),
we have $e_2 \in X$. By (\ref{eq:XX}), all $x_j$ are ears of $S$
for $j = 1 , 2, \ldots, k$. It follows from (\ref{eq:X}) that $S$
has exactly $k$ ears in $U$. Therefore, $S$ is a $2k$-sun. We have
proved (\ref{eq:sun}).

We continue with the proof of the Claim (and the Theorem).
Consider the $k$ ears of $S$ that belong to $U$. Since each $V_i$
is a clique, it contains at most one ear. So, there are $k$ sets
$V_i$ containing an ear of $S$. Let these sets be $V_1, V_2,
\ldots, V_k$.  Clearly, in $G$, the vertices $v_1, v_2, \ldots ,
v_k$ form a stable set. $\Box$

\noindent
{\it Proof of Theorem \ref{thm:sun2}.} We will use the notation
defined in the proof of Theorem \ref{thm:sun} with $G$ being a
triangle-free graph. We only need prove the graph $f(G,k)$ does
not contain a $t$-antihole with $t \geq 7$. We will prove by
contradiction. Suppose $f(G,k)$ contains a $t$-antihole $A$ with
vertices $a_1, a_2, \ldots, a_t$ with $t \geq 7$ such that $a_i$
misses $a_{i+1}$ with the subscripts taken modulo $k$. Since the
vertices in $X$ have degree two, none of them can belong to $A$.
Since each $V_i$ is a clique,
\begin{eqnarray}
\mbox{no two consecutive vertices of $A$ can belong to the same
$V_i$.}\label{eq:Vi}
\end{eqnarray}
Similarly,
\begin{eqnarray}
\mbox{no two consecutive vertices of $A$ can belong to
$W$.}\label{eq:W}
\end{eqnarray}
Now, we claim that
\begin{eqnarray}
\mbox{one of $a_i, a_{i+1}$ must lie in $W$ for all
$i$.}\label{eq:edge}
\end{eqnarray}
Suppose (\ref{eq:edge}) is false for $a_i$. For simplicity, we may
assume $i=1$, and so we have $a_1, a_2 \in U$. By (\ref{eq:Vi}),
we may assume $a_1 \in V_1, a_2 \in V_2$. Clearly, we have $a_t
\not\in V_1$.

Suppose $a_t \in V_2$. Then $a_3$ has to be in $W$, for otherwise
$a_3$ lies in some $V_j$ and so it misses $a_t$ (since it misses
$a_2$) implying $t = 4$, a contradiction. By symmetry, we have
$a_{t-1} \in W$. Since $a_1$ sees $a_3$, and $a_{t-1}$ is a common
neighbour of $a_1$ and $a_2$, Observation \ref{obs:VW} implies
that $a_2$ sees $a_3$, a contradiction to the definition of $A$.
So, we have $a_t \not\in V_2$.

Suppose $a_t \in W$. By (\ref{eq:W}), we have $a_{t-1} \in V_j$.
If $j=2$ then $a_1$ misses $a_{t-1}$, a contradiction to the
definition of $A$. If $j=1$ then $a_2$ misses $a_{t-1}$ implying
$t=4$, a contradiction. So, we may assume $a_{t-1} \in V_3$. Let
$j \in \{t-2, t-3\}$.  If $a_j \in W$ then since $a_2$ sees $a_t$,
Observation \ref{obs:VW} with $z=a_j, x=a_2, y=a_1$ implies $a_1$
sees $a_t$, a contradiction to the definition of $A$. So, we have
$a_{t-2} \in V_m$ for some $m$, and  $a_{t-3} \in V_p$ for some
$p$. Since $a_{t-2}$ misses $a_{t-1}$, we have $a_{t-2} \not\in
V_1 \cup V_2 \cup V_3$. So, we may assume $m=4$. We have $a_{t-3}
\in V_2 \cup V_3$, for otherwise the three vertices $a_{t-3},
a_{t-1}, a_2$ contradict Observation \ref{obs:triangle}. Since
$a_{t-2}$ sees $a_2$, $a_{t-2}$ sees all of $V_2$. Thus, we have
$a_{t-3} \not\in V_2$, and so $a_{t-3} \in V_3$.
Since $t \geq 7$, the vertex $a_{t-4}$ exists. Since $a_{t-4}$
misses $a_{t-3}$ but sees $a_{t-1}$, $a_{t-4}$ is not in $U$; so
we have $a_{t-4} \in W$. Observation \ref{obs:VW} with $z=
a_{t-4}, x= a_{t-1}, y= a_{t-2}$ implies $a_{t-1}$ sees $a_t$, a
contradiction to the definition of $A$.

Thus, $a_t$ belongs to some $V_j$ which is distinct from $V_1,
V_2$. It follows from symmetry and the definition of $f(G,k)$ that
$a_3, a_{t-1}$ also belong to distinct $V_i$. Now, the three
vertices $a_{t-1}, a_1, a_3$ contradict Observation
\ref{obs:triangle}. So, (\ref{eq:edge}) holds.

From (\ref{eq:W}) and (\ref{eq:edge}), we may assume without loss
of generality that $a_i \in U$ whenever $i$ is odd, and $a_i \in
W$ whenever $i$ is even. In particular, $t$ is even and at least
eight. The definition of $A$ implies that $a_1$ sees $a_4, a_6$.
Thus, we have $\{a_4, a_6\} = \{u_i, w_i\}$ for some $i$. The
definition of $f(G,k)$ means that every vertex of $U$ either sees
both $a_4, a_6$ or misses both of them. But $a_3$ misses $a_4$ and
sees $a_6$, a contradiction. $\Box$

\noindent {\it Proof of Theorem \ref{thm:ksun}.} We will reduce
$k$-CLIQUE to $k$-SUN. Let $G,k$ be an instance of $k$-CLIQUE. We
may assume $k \geq 4$. Construct a graph $h(G)$ from $G$ by adding
a vertex $v(a,b)$ for each edge $ab$ of $G$, and joining $v(a,b)$
to $a$ and $b$ by an edge of $h(G)$. Let $Y$ be the set of
vertices $v(a,b)$. It is easy to see that if $G$ has a clique $K$
on $k$ vertices then $h(G)$ has a $k$-sun induced by $K$ and some
$k$ vertices in $Y$. If $h(G)$ has a $k$-sun $(c_1, \ldots, c_k,
e_1, \ldots, e_k)$, then since the vertices in $Y$ have degree
two, none of them can be a vertex $c_i$; thus, the vertices $c_1,
\ldots, c_k$ induce a clique on $k$ vertices in $G$. $\Box$

\end{document}